# Nitrogen in Silicon for Room Temperature Single Electron Tunneling Devices

Pooja Yadav[1], Hemant Arora[1], and Arup Samanta[1,2*]

[1]Department of Physics, Indian Institute of Technology Roorkee, Roorkee-247667, Uttarakhand, India
[2]Centre of Nanotechnology, Indian Institute of Technology Roorkee, Roorkee-247667, Uttarakhand, India
*Corresponding author: arup.samanta@ph.iitr.ac.in

Single electron transistor (SET) is an advanced tool to exploit in quantum devices. Working of such devices at room-temperature is essential for practical utilization. Dopant based single-electron devices are well studied at low-temperature although a few devices are developed for high-temperature operation with certain limitations. Here, we propose and theoretically exhibit that nitrogen (N) donor in silicon is an important candidate for effective designing of such devices. Theoretical calculation of density-of-states using semi-empirical DFT method indicates that N-donor in silicon has deep ground state compared to a phosphorus (P) donor. N-donor spectrum is explored in nano-silicon along with the P-donor. Comparative data of Bohr radius of N-donor and P-donor is also reported. The simulated current-voltage characteristics confirm that N-doped device is better suited for SET operation at room-temperature.

In the past two decades, a significant effort had been given to develop single electron tunneling devices using dopant atom for solid state-based quantum architecture by utilizing and manipulating the charge and spin degrees of freedom using external fields.[1–4] Dopant based SET devices are key components in quantum device applications e.g., quantum bit[1], memories[5], single electron pump[6], single charge sensing[7], single photon detectors[8] etc. Extensive work has been put forward on a few dopant based devices where single electron tunneling effect are observed at low temperatures.[9–14] However, the room temperature operation is highly needed for practical utilization of such devices. The reason being the low barrier height and charging energy due to shallow nature of dopants e.g., arsenic (As) phosphorous (P) as donor and boron (B) as acceptor in silicon. Some results on high temperature operation of these devices due to quantum confinement, dielectric confinement, and donor's cluster formation have also been reported.[15,16]

To improve the operating temperature of such SET, alternative dopants are also proposed for the utilization of double donors like tellurium (Te), selenium (Se), and sulfur (S), which have one order higher binding energy than shallow donors.[17] Recently, the spin relaxation and donor-acceptor recombination of Se$^+$ in $^{28}$Si was reported.[18] However, use of double donors needs control of several associated effect due to the existence of multiple electrons in the donor side. Erbium being a deep donor has also been studied for spin-based devices.[19] High temperature operations of the germanium vacancy complexes based devices have recently been proposed and demonstrated.[20,21] Room temperature operation of deep donor pair of Al-N in silicon is reported in TFET configuration.[22] However, the use of a single deep dopant is better for quantum operation due to the minimization of decoherence path.

Here, we propose an alternative quantum architecture for such approach using nitrogen (N) donor in FinFET/SOI-FET configuration for high temperature operation and also explored the N-donor spectrum in bulk-silicon, nano-silicon, and in device configuration. Nitrogen energy level is ~190 meV below the conduction band of bulk silicon, which can be a suitable candidate for high temperature single electron tunneling devices.[23] This means the discrete states of N-donor can be preserved at room temperature—by suppressing the crosstalk with external environment and saving the electron states from thermal demons. Accounting the natural abundance of Si$^{28}$ isotope as a host for nitrogen donor, electron and spin states can also be conserved for a longer time considering the deep donor energy levels.[24] In addition, this system could also be an important candidate for nuclear spin qubit by analogy with P-donor in silicon, since the different isotopic nuclear spin of N-donor and the zero isotopic spin of the silicon host medium.[1]

In this letter, undoped, phosphorous doped and nitrogen doped hydrogenated silicon nanowire (H:SiNW) in normal and SET configurations along with the bulk structure are designed and the comparative analyses using semi-empirical density functional theory (DFT) calculation for these configurations were performed. The projected density of states (PDOS), total density of states (TDOS), local density of states (LDOS) along with the current-voltage ($I_{DS}$-$V_G$) characteristics are studied. Single electron tunneling transport through the discrete energy states are investigated for both P-doped and N-doped devices at low and room temperatures.

The numerical calculations are performed on Quantum ATK software[25] based on density functional theory. Semiempirical extended-Hückel[26] method is used for these calculations at 300 K. The mesh cuttoff was set at 20 Hartree. The k grid 1×1×174 is used. The atomistic structure of a silicon nanowire of length 32.4 Å and diameter 10 Å is constructed along [100] direction. The surface dangling bonds are passivated by the hydrogen atoms. The schematic and atomistic design of the structures are presented in Fig. 1(a-b). An in-built optimizer is utilized for structure optimization with force constant 0.008 eV$^{-1}$. This structure is used to calculate the TDOS and PDOS spectra of H:SiNW. In addition, we examined the density of states profile of undoped, P-doped, and N-doped bulk silicon used with 4x4x4 fcc supercell (512 Si lattice sites). The density mesh cut-off was set to 20 Hartree with 4x4x4 k-point set proposed by Makov, Shah, and Payn.[27]

To study donor-based transistor devices, the SiNW transistors having diameter of 10 Å and length of 54 Å are also structured by gate all around assembly along with tunnel coupling of the center dopant with source and drain reservoirs as shown in Fig 1(c). In these devices, the gate covers the donor position of the channel of the SET devices. Vacuum is used as a gate dielectric for the present configurations. The source and drain leads are heavily doped with n-type dopant ($1\times10^{21}$ cm$^{-3}$). $I_{DS}$-$V_G$ characteristics of these devices are calculated using the non-equilibrium Green's function (NEGF) method with Landauer formalism.[28]

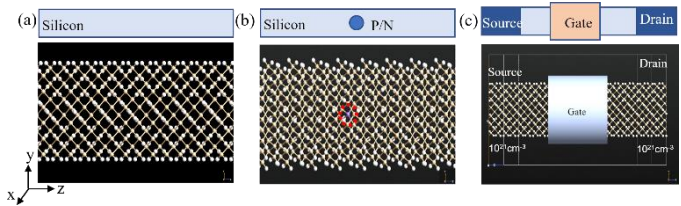

Figure 1: (a) Quantum ATK simulated structure for (a) H:SiNW, (b) substitutional P/N-doped (encircled by red dotted line) H:SiNW, and (c) P/N-doped H:SiNW in transistor configuration with gate terminal and source drain terminals. Top Panel: schematic of (a-c) structures. SiNW radius is in x-y dimension and z dimension is along the length axis.

Initially the TDOS spectrum of undoped, P-doped and N-doped bulk system are studied in Fig. 2 (a-c). The undoped bulk silicon shows a band gap of 0.6 eV as presented in Fig. 2a, consistent with previous reports.[29] The P-doped and N-doped bulk Si systems are studied subsequently. Here, the Fermi energy is set at 0 eV for all results. The donor energy state is found at $E-E_F$ = 0 eV for both the cases, which are the ground states of the respective donors. The separation between the donor ground state (GS) and next conducting state depicts the shallow and deep nature of the P-donor and N-donor, respectively. In the TDOS spectrum of P-donor in Fig. 2 (b), the mixed states are present at $E-E_F$=0 eV, and donor states cannot be distinctly separated due to thermal smearing. In the TDOS spectrum of N-doped bulk silicon as shown in Fig. 2(c), the donor state lies at 0 eV and the next conducting state at 184 meV. Since the donor states are merged in the conduction band of bulk structure, it is difficult to separate further donor states. Hence, the separation of the GS of N-donor with conduction band edge (CBE) is 184 meV, consistent with experimentally reported value.[30]

Now to get a full insight of the donor states and a detailed analysis of donor behavior in nano-silicon devices, we investigated the same analysis in H:SiNW systems. From now onwards, all the calculations are performed on H:SiNW. The TDOS spectra of undoped, P-doped, and N-doped H:SiNW is calculated and presented in Fig. 2(d-f), respectively. In the TDOS spectrum of undoped H:SiNW as shown in Fig. 2(d), the separation between the conduction-band and valence-band edges is very high (~1.7 eV), indicating a wide band gap due to the confinement effect.

Modification in the TDOS features is observed in both P-doped and N-doped systems compared to undoped H:SiNW as presented in Fig. 2(e) and 2(f), respectively. Additional states within the bandgap are observed in doped structures. To probe the TDOS spectra in details, the PDOS spectra are studied to identify and locate the states. It is well documented that the donor energy level in silicon splits into three energy groups due to valley degeneracy: one $A_1$ state (lowest energy state), one threefold $T_2$ state (1st excited state), and one twofold E state (2nd excited state).[31,32] The $A_1$ orbital shows s-like symmetry on the dopant location of nanostructure while the $T_2$ and E orbitals show a node at this position.[33] The $A_1$, $T_2$, and E states are identified for both P-donor and N-donor in this study in comparison with the literature.[34]

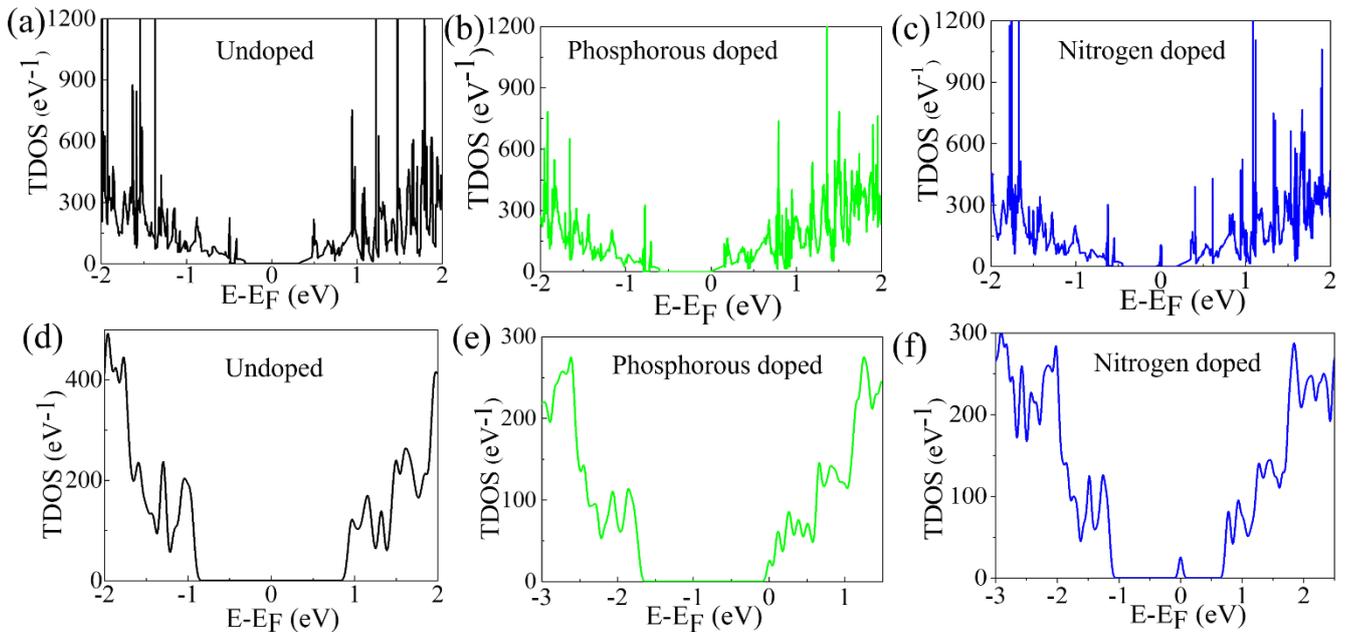

Figure 2: (a-c) The TDOS spectra of undoped, P-doped, and N-doped bulk silicon. (d-f) The TDOS spectra of undoped, P-doped, and N-doped H:SiNW.

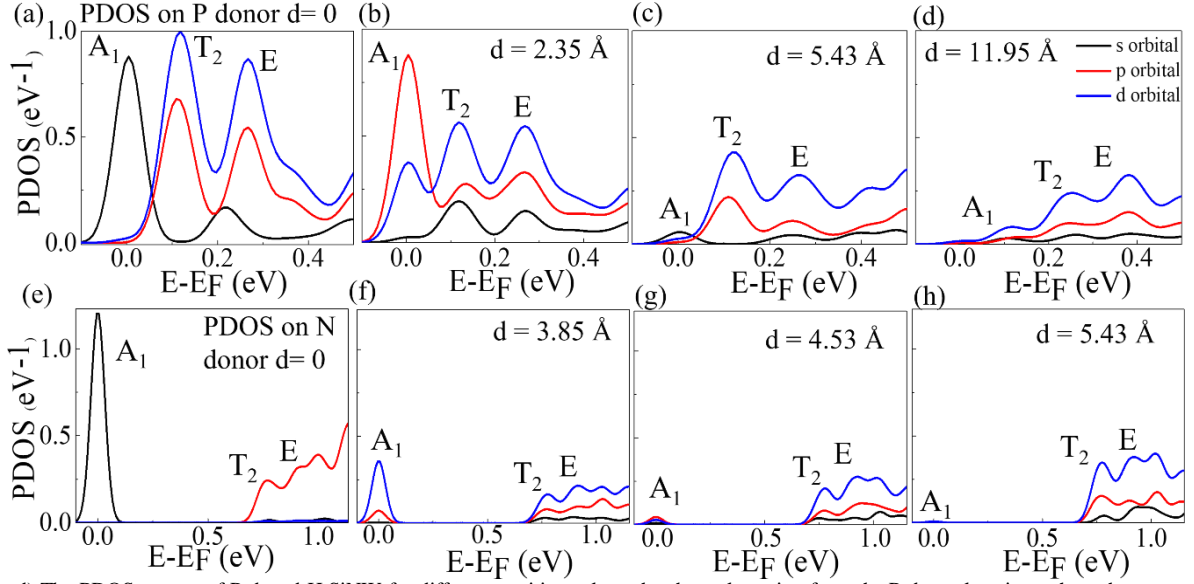

Figure 3. (a-d) The PDOS spectra of P-doped H:SiNW for different positions along the channel starting from the P-donor location, where the states $A_1$, $T_2$, and E are identified. (e-h) Same spectra for N-donor atom.

We focused on the position dependent PDOS of the P-donor in Fig. 3(a-d). The PDOS components of s, p, and d orbitals of the P-donor on itself i.e., d=0 exhibit that $A_1$, $T_2$, and E energy levels exist at this location, where $A_1$ state is the lowest energy state and it is mainly coming from s-orbital of the P-donor.

The $T_2$ and E energy levels are the mixture of s, p, and d orbitals. The PDOS spectrum of P-donor on different silicon locations (d starts from the donor location and extends towards drain reservoir side) are investigated in Figs. 3(b-d). The existence of the PDOS on the silicon position confirms that the hybridization is present between the electronic states of P and Si atoms. We mainly focus on the ground state (i.e., $A_1$) of the P-donor, which diminishes as we move away from the donor location. The localization of the $A_1$ state can be correlated with the spreading of this state. Such behavior can be linked to the Bohr radius of the ground state, which is the distance from where the $A_1$ state originates and where the $A_1$ state dies out. The extinction of the $A_1$ state is the end of this state and its length from the donor location is d =16.29 Å. Hence, the Bohr radius of P-donor in such system is ~ 16.29 Å, which is far smaller than the bulk case (~25 Å).[31] The reduction in the Bohr radius happens due to the quantum confinement effect. We study the PDOS spectra of N-donor in H:SiNW as shown in Fig. 3(e-h) and found a similar trend of features as we shift far from the donor location. The PDOS results for $A_1$ state reduce to negligible value at d=5.43 Å, which signifies the ground state is highly concentrated and the Bohr radius is ~5.43 Å. We also observe that $A_1$ state is largely separated from any other states of this configuration. The difference between N-donor's ground state and the first excited state (ES) is 770 meV while that for P-donor is 118 meV. This is consistent with the highly localized ground state of N-donor. Such highly localized and deep ground state is the pathway for designing quantum devices for room temperature.

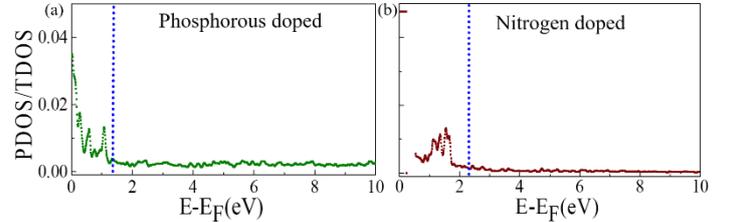

Figure. 4: (a-b) PDOS/TDOS spectra of a P-doped and N-doped silicon nanowire, respectively.

To calculate the binding energy of donor electron ($E_b$), we estimated the lowest effective conductive state of doped H:SiNW following Anh et al.[33] Since the donor electronic states hybridize with silicon electronic states, the conductive state changes in comparison with undoped system. The $E_b$ is estimated by taking the ratio of PDOS/TDOS, which signifies the relative weight of donor states against the states of the whole system. Figures 4(a) and (b) show the PDOS/TDOS spectra for P-donor and N-donor, respectively. At lower energy range, the contribution of the donor is dominated, while the contribution is trivial at higher energy range (above blue dashed line) due to the inclusion of more number of silicon atoms. The binding energy is calculated from the point where this ratio saturates close to 0. The calculated $E_b$ values for P-donor and N-donor are 1.36 eV and 2.40 eV, respectively.

In order to understand the behavior of undoped and doped devices, LDOS spectra are simulated along with the current vs gate voltage ($I_{DS}$-$V_G$) characteristics for different source to drain voltage ($V_{DS}$) using the transistor device parameters described in Fig. 1(c). The LDOS spectrum of the undoped H:SiNW transistor is obtained where no states are observed below the conduction band edge as shown in Fig. 5(a). In Fig. 5(b) and 5(c), extra energy states are observed in the LDOS data due to substitutionally doped P-donor and N-donor, respectively.

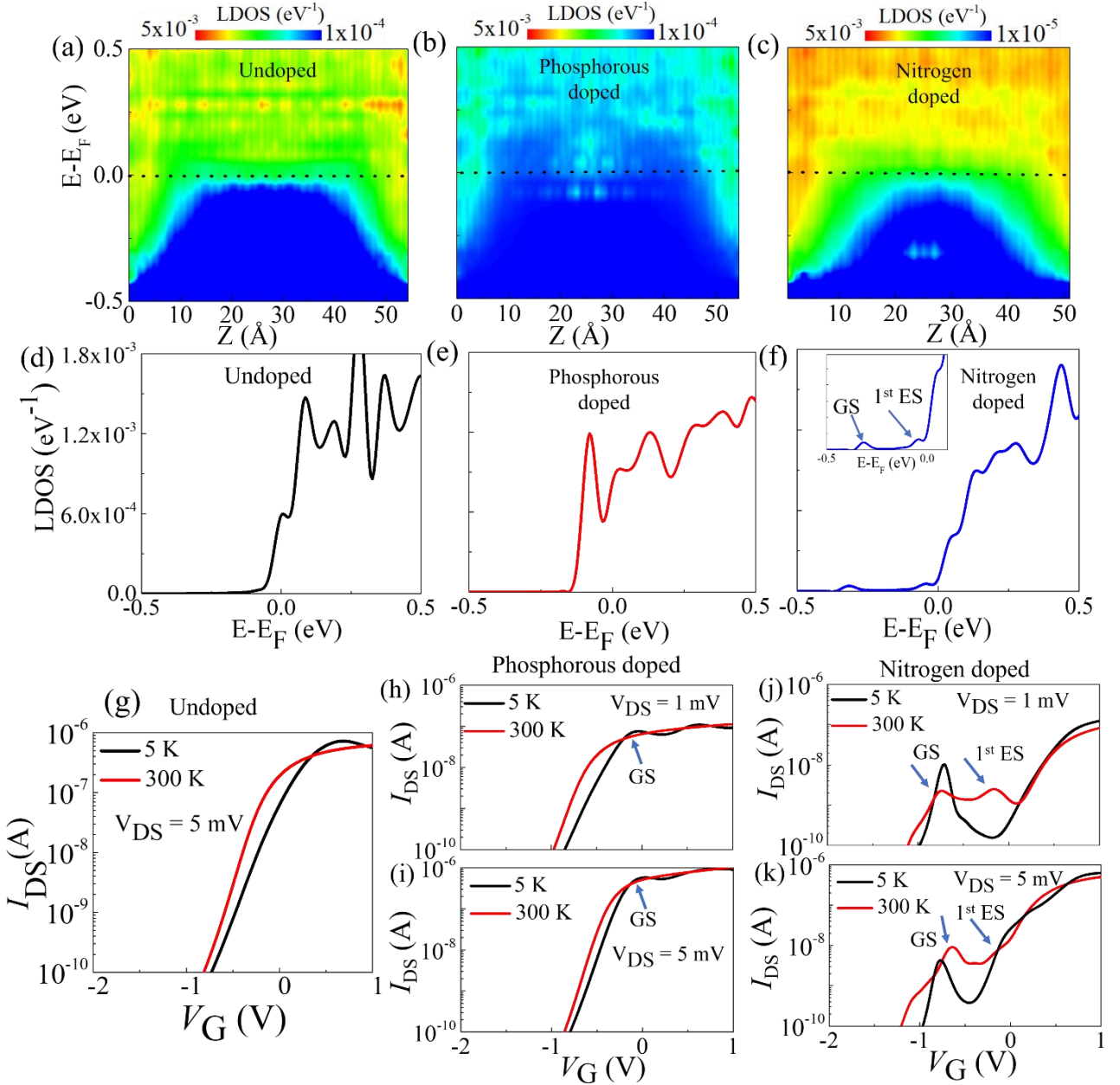

Figure. 5: (a-c) LDOS spectra of an undoped, P-doped, and N-doped silicon nanowire transistor, respectively. (d-f) LDOS spectrum at z = 26.93 Å for undoped, P-doped, and N-doped silicon nanowire transistor, respectively. Inset of Fig. 5(f) indicates the zoom in spectrum, exhibiting the GS and and 1st ES of N-donor. Simulated $I_{DS}$-$V_G$ characteristics for (g) undoped, (h-i) P-doped, and (j-k) N-doped devices.

The lowest energy state in the channel region of P-doped device is found at 80 meV from the next conductive state, which is low compared to non-device configuration (118 meV). This is most likely due to the effect of metal gate and closely spaced heavily doped leads. In LDOS graph of N-doped device, a very deep energy state is observed at 322 meV below the nearest conductive state. To emphasize more on the donor states obtained in LDOS spectra, the same information of the LDOS spectra is mapped along the vertical axis in Fig. 5(d-f). The LDOS spectra along the Z axis at z = 26.93 Å are presented for all three device structures. No LDOS is observed below CBE for undoped device as shown in Fig. 5(d). A single LDOS peak in P-doped SiNW at 80 meV is observed in Fig. 5(e), indicating the GS of P-donor in the device configuration. A highly localized LDOS is also perceived at 322 meV from the CBE in Fig. 5(f), providing the information of the GS of N-dopant in the device configuration. In addition, we also observed 1st ES of N-donor at 54 meV below the CBE.

The $I_{DS}$-$V_G$ transport characteristics for all three devices are simulated at the different bias voltages for T= 5 K and 300 K, as presented in Figs. 5(g-k). In the undoped device with $V_{DS}$= 5 mV, the characteristics are typical FET devices, which is consistent with the LDOS spectrum of this device configuration. The weak modulation in $I_{DS}$ for T = 5 K is observed at ~ $V_G$=0.5 mV in Fig. 5(g), which is due to the transport through the discrete states in the CB of the undoped H:SiNW. It is also observed that the subthreshold slopes of $I_{DS}$-$V_G$ curves at 5 K and 300 K are almost same that may be due to the effect of heavy doping ($10^{21}$ cm$^{-3}$) of the leads along with the nano-channel effect.

The transport characteristics for P-doped device are presented in Fig. 5(h)-(i) for $V_{DS}$= 1 mV and 5 mV, respectively,

showing a strong single electron tunneling current peak before the on-set of FET current at T=5 K due to the existence of localized state for P-donor within the channel window. An additional weak current peak is also preserved due to the transport through the next discrete conducting state. However, this SET peak dies out at 300 K due to the shallow nature of such dopant.

The $I_{DS}$-$V_G$ characteristics for N-doped device are presented in Fig. 5(j-k) for $V_{DS}$= 1 mV and $V_{DS}$ = 5 mV, respectively. For $V_{DS}$= 1 mV, a very strong tunneling peak is observed via the GS of N-donor. However, the GS and 1st ES prominently participated in the transport characteristics at T=300 K. The excited state features got observed at high temperature due to higher tunnel rate compared to 5 K. For $V_{DS}$= 5 mV and T=5 K, the low temperature behavior of the device is similar to $V_{DS}$= 1 mV with lower separation from the FET current, and this current peak is also strongly sustained at T=300 K. Along with the transport through the GS at 300 K, a small hump corresponding to 1st excited state is also noted. Observation of a single electron current peak at room temperature is happened due to deep nature of N-donor. Such deep donor state is capable of holding the quantum information and is less fragile to environment fluctuations compared to shallow donor. Silicon being one of the most promising semiconductor for quantum information devices due to large charge and spin coherence times,[34,35,36] donor state of nitrogen in silicon gives an opportunity to utilize it for better performances. Hence, to design single electron transistor with a single donor atom for room temperature operation, N-donor in silicon could be one of the best systems. We are working in such direction for the experimental realization of such devices.

Donor energy spectrum of nitrogen in hydrogenated-silicon nanowire is theoretically investigated in comparison with phosphorous and undoped system. We observed that binding energy of N-donor ground state is large compared to the P-donor. It is also highly localized and separation of ground state is also very large compared to 1st excited state. In quantum devices where perseverance of quantum states is important, this advantage can be taken from a natural system with deep N-donor states. Simulated LDOS and $I_{DS}$-$V_G$ characteristics of the devices also suggest in favor for N-dopant being the potential candidate for single electron tunneling devices for room temperature operation, which further infers this system could be useful for designing practical quantum architecture.

The authors are thankful to Prof. Sparsh Mittal for the computational resources under the project ECR/2017/000622. The work is partially supported by DST-SERB (Project no: ECR/2017/001050) and IIT Roorkee (Project no: FIG-100778-PHY), India. P.Y. and H.A. thank to Ministry of Education and UGC, India, respectively for the research scholarship.